\documentclass[pra,twocolumn]{revtex4}
\usepackage{graphicx}

\begin{document}
\title{Quantum dynamics of two capacitively coupled superconducting islands via Josephson junctions}
\author{ Mou Yang and Le-Man Kuang\footnote{Corresponding author.}\footnote{
Email: lmkuang@hunnu.edu.cn}}
\address{Department of Physics, Human Normal University, Changsha 410081, China\ }

\begin{abstract}
In this paper, we consider a system consisting of two capacitively
coupled superconducting islands via Josephson junctions. We show
that it can be reduced to two coupling harmonic oscillators under
certain conditions, and solved exactly in terms of a displacing
transformation, a beam-splitter-like transformation, and a
squeezing transformation. It is found that the system evolves by a
rotated-squeezed-coherent state when the system is initially in a
coherent state. Quantum dynamics of the Cooper pairs in the two
superconducting islands is investigated. It is shown that the
number of the Cooper pairs in the two islands evolves
periodically.

\vspace{1cm}
\noindent PACS number(s): 74.50+r, 73.23.Hk, 03.67.-a

\end{abstract}

\maketitle

%%%%%%%%%%%%%%%%%%%%%%%%%
%\section{Introduction}
%%%%%%%%%%%%%%%%%%%%%%%%%
Recent experiments \cite{nak,vio} have demonstrated the
possibility of controlling or manipulating macroscopic quantum
states in a single-Cooper-pair box, which consists of a small
capacitance supuerconducting island coupled to a bulk
superconductor via Josephson tunneling. Due to the small
capacitance, there can be only zero or one excess Cooper pair on
the island, which leads to an effective two-level system at
appropriate values of external bias voltage. The possible use of
these two-level systems as building blocks of  quantum computer
had been suggested already prior to the experimental work
\cite{shn,mak}. However, it is clear that realizing such a quantum
computer even with a modest number of qubits will prove
exceedingly difficult and progress will be made only in small
steps. A first step in this direction will consist in the coupling
of two such Cooper-pair boxes.

Marquardt and Bruder  \cite{mar} studied the quantum dynamics of a
system which consists of a Cooper-pair box and a capacitively
coupling large superconducting island, predicted generation of
mesoscopically distinct quantum states. The large superconducting
island can be described by a harmonic oscillator since it has a
comparatively small charging energy, then we have $E_J/E_C$
approache the infinity, where $E_J$ and $E_C$ denote the Josephson
coupling energy and the charging energy of the large island,
respectively. The coupling system of a Cooper-pair box and a large
superconducting island is analogous to a system which consists of
a two-level atom interacting with a single mode of the quantized
electromagnetical field.

On the other hand, recently much attention has been paid to
continuous-variable quantum information processing
\cite{jeo,enk,got,ban,bra,fur,zhou,cai} which is based on harmonic
oscillators in infinite dimensional Hilbert spaces. Therefore, it
is is of significance to study such systems of capacitively
coupling large superconducting islands via Josephson junctions in
order to explore continuous-variable quantum information
processing realizations in solid-state systems. In this paper, we
consider theoretically another possibility: two large
superconducting islands are capacitively coupled via two Josephson
junctions. Each island may be described by a harmonic oscillator,
thus the system is reduced to a two-coupling-harmonic-oscillator
system.  We shall give an analytical solution of this system and
investigate the quantum dynamics of the Cooper pairs in the
islands.

Consider a system which consists of two capacitively coupled superconducting
islands described in Fig. 1 where $C_{J_i}$ and $C_{g_i}$ are the capacitances
of $i$-th Josephson junction and $i$-th gate capacitor, respectively, $C_c$ is
the coupling capacitance, and $V_i$  denotes gate voltage. The total energy of
this system is the sum of the charging energy and the Josephson coupling energy.
The charging energy is given by
\begin{equation}
\label{1}
E_{ch}=\sum^2_{i=1}\left[ \frac{Q^2_{J_i}}{2C_{J_i}} + \frac{Q^2_{g_i}}{2C_{g_i}}
+ V_i(Q_i-Q_{J_i})\right]+ \frac{Q^2_{c}}{2C_{c}},
\end{equation}
where $Q_i$ is the total charge on the $i$-th island, $Q_{c}$ the charge on the
coupling capacitor, and $Q_{J_i}$ and $Q_{g_i}$ denote the charges on $i$-th Josephson
junction and $i$-th gate capacitor, respectively.

The charging energy  can be expressed as a function of the total charges on the two
superconducting islands $Q_1$ and $Q_2$ alone and the circuit parameters.
If let $N_i=Q_i/2e$ denote the number of the Cooper pairs on the $i$-th island with
$e$ being the charge of the electron, then the charging energy can written as
 \begin{eqnarray}
\label{2}
E_{ch}&=&\sum^2_{i=1} E_{C_i} (N_i-n_{g_i})^2 + E_{12}(N_1-n_{g_1})(N_2-n_{g_2}) \nonumber \\
& & + N_i-\textrm{independent terms},
\end{eqnarray}
where we have introduced the following notations
\begin{eqnarray}
\label{3}
E_{12}&=&\frac{(2e)^2C_C}{C_{t_1}C_{t_2}-C^2_C}, \\
\label{4}
E_{C_1}&=&\frac{(2e)^2C_{t_2}}{2(C_{t_1}C_{t_2}-C_C^2)}, \\
\label{4'} E_{C_2}&=&\frac{(2e)^2C_{t_1}}{2(C_{t_1}C_{t_2}-C_C^2)},\\
C_{t_i}&=&C_{J_i}+C_{g_i}+C_C,
\end{eqnarray}
and the offsets introduced by the gates are given by
\begin{eqnarray}
\label{5} n_{g_i}=\frac1{2e} \left[C_{g_i}V_i-(-1)^i
C_C(V_2-V_1)\right].
\end{eqnarray}
The Josephson coupling energy of the two superconducting islands
is given by
\begin{equation}
\label{11} E_{Jos}=-\sum^2_{i=1}E_{J_i}\cos\phi_i.
\end{equation}

Quantum mechanically, the number of the Cooper pairs and the
Josephson phase are regarded as inter-conjugate operators, and
satisfy the boson communication relation $[\hat{\phi_j},
\hat{N_k}]=i\delta_{jk}$. Then we get the Hamiltonian
\begin{eqnarray}
\label{12}
\hat{H}&=& \sum^2_{i=1} \left[E_{C_i}(\hat{N}_i -n_{g_i})^2 - E_{J_i}\cos\hat{\phi}_i\right] \nonumber \\
& & +E_{12}(\hat{N}_1 -n_{g_1})(\hat{N}_2 -n_{g_2}),
\end{eqnarray}
where we have discard the constant terms.

For a large superconducting island, the Josephson coupling energy
is much larger than the charging energy, i.e., $E_{J_i}/E_{C_i}
\to \infty$. In this case the approximation of a harmonic
oscillator is valid, then one can replace $\cos\hat{\phi}_i$ term
by the parabolic potential $1-\hat{\phi}^2_i/2$. Hence  the
Hamiltonian (\ref{12}) can be understood as that of two coupled
harmonic oscillators. After making the following displacing
transformation
\begin{equation}
\label{13}
\hat{d}_1(n_{g_1}, n_{g_2})=\exp\left[i(\hat{\phi}_1 n_{g_1}+ \hat{\phi}_2n_{g_2})\right],
\end{equation}
the Hamiltonian (\ref{12}) can be expressed in terms of boson
annihilation and creation operators as follows
\begin{eqnarray}
\label{14}
\hat{H}_1&=&\sum^2_{i=1}\hbar\omega_i\hat{a}^{\dagger}_i\hat{a}_i +
\lambda\hbar(\hat{a}^{\dagger}_1\hat{a}_2 + \hat{a}^{\dagger}_2\hat{a}_1) \nonumber \\
& &- \lambda\hbar(\hat{a}^{\dagger}_1\hat{a}^{\dagger}_2 + \hat{a}_1\hat{a}_2),
\end{eqnarray}
where we have used the following decompositions
\begin{eqnarray}
\label{15}
\hat{\phi}_i&=&\left(\frac{E_{C_i}}{2E_{J_i}}\right)^{1/4}
\left(\hat{a}^{\dagger}_i + \hat{a}_i\right), \hspace{1cm} (i=1, 2), \\
\label{16}
\hat{N}_i&=&i\left(\frac{E_{J_i}}{8E_{C_i}}\right)^{1/4}\left(\hat{a}^{\dagger}_i
- \hat{a}_i\right), \hspace{0.8cm} (i=1, 2),
\end{eqnarray}
and the free-evolution frequencies and the coupling constant are given, respectively, by
\begin{eqnarray}
\label{17}
\omega_i&=&\frac 1{\hbar}\sqrt{2E_{C_i}E_{J_i}},   \hspace{1cm} (i=1, 2), \\
\label{18}
\lambda&=&\frac{E_{12}}2\left(\frac{E_{J_1}E_{J_2}}{4E_{C_1}E_{C_2}}\right)^{1/4}.
\end{eqnarray}

Then the displacement operator (\ref{13}) can be rewritten in
terms of the annihilation and creation operators as
\begin{eqnarray}
\label{19}
\hat{d}_1(n_{g_1}, n_{g_2})&=&\prod^2_{i=1}\exp\left[\alpha_{0i}\hat{a}^{\dagger}_i
-\alpha^*_{0i}\hat{a}_i\right] \nonumber \\
&\equiv& \hat{D}(\alpha_{01}, \alpha_{02}),
\end{eqnarray}
where $\alpha_{01}$ and $\alpha_{02}$ are given by
\begin{equation}
\label{20}
\alpha_{01}=in_{g_1}\left(\frac{E_{C_1}}{2E_{J_1}}\right)^{1/4},
\alpha_{02}=in_{g_2}\left(\frac{E_{C_2}}{2E_{J_2}}\right)^{1/4}.
\end{equation}

%%%%%%%%%%%%%%%%%%%%%%%%%%%%%%%%%%%%%%%%%%%%%%%%%%%%%%%%%%%%%%%%%%%%%%%%%%%%%%%%%%%%%%%%
\begin{figure}
\begin{center}
\includegraphics[width=4.6in,height=3.5in]{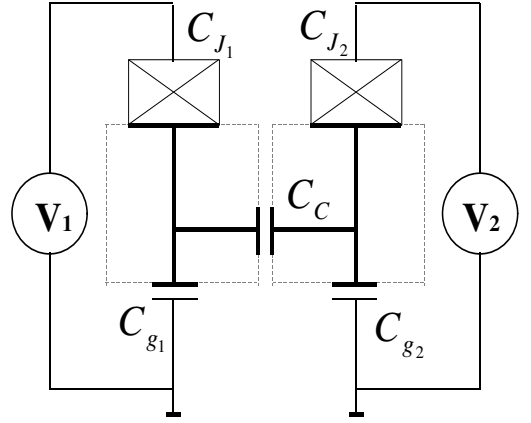}
\end{center}
\caption{Circuit diagram for the considered system. The two
superconducting islands appearing as boxes corresponding to the
regions marked by dashed rectangles. The junction and gate
capacitances are $C_{J_i}$ and $C_{g_i}$ for $i$-island,
respectively. The coupling capacitance is denoted by $C_C$ and the
gate voltages are $V_1$ and $V_2$.}
\end{figure}
%%%%%%%%%%%%%%%%%%%%%%%%%%%%%%%%%%%%%%%%%%%%%%%%%%%%%%%%%%%%%%%%%%%%%%%%%%%%%%%%%%%%%%%%

For the sake of simplification, we consider the symmetric
situation where $\omega_1=\omega_2\equiv\omega$ since $C_1=C_2$
and $C_{J_1}=C_{J_2}$. In this case, the cross term in the
Hamiltonian (\ref{14}) can be gotten rid of by the following
beam-splitter-like transformation over
\begin{equation}
\label{21}
\hat{B}(\varphi)=\exp\left[\varphi(\hat{a}^{\dagger}_1\hat{a}_2 - \hat{a}^{\dagger}_2\hat{a}_1)\right].
\end{equation}

From (\ref{14}) and (\ref{21}) we obtain the transformed Hamiltonian
\begin{eqnarray}
\label{22}
\hat{H}_2&=& \hat{B}^{\dagger}\left(\frac{\pi}{4}\right)\hat{H}_1
\hat{B}\left(\frac{\pi}{4}\right) \nonumber \\
&=& \hbar\omega'_1\hat{a}^{\dagger}_1\hat{a}_1 +
     \hbar\omega'_2\hat{a}^{\dagger}_2\hat{a}_2 \nonumber \\
& &+ \frac{\hbar\lambda}{2}\left(\hat{a}^{\dagger 2}_1 + \hat{a}^2_1\right)
   - \frac{\hbar\lambda}{2}\left(\hat{a}^{\dagger 2}_2 + \hat{a}^2_2\right),
\end{eqnarray}
where we have let $\omega'_1=\omega_1-\lambda$ and
$\omega'_2=\omega_2+\lambda$. In the derivation of the Hamiltonian
(\ref{22}) we have used the following formula
\begin{eqnarray}
\label{23}
\hat{B}^{\dagger}(\varphi)\hat{a}_1\hat{B}(\varphi)&=&
\hat{a}_1\cos\varphi + \hat{a}_2\sin\varphi, \\
\label{24}
\hat{B}^{\dagger}(\varphi)\hat{a}_2\hat{B}(\varphi)&=&
\hat{a}_2\cos\varphi - \hat{a}_1\sin\varphi.
\end{eqnarray}

Finally, we make a squeezing transformation over the Hamiltonian
(\ref{22}) to get that
\begin{equation}
\label{25}
\hat{H}_3= \hat{S}^{\dagger}\left(\xi_1, \xi_2\right)\hat{H}_2
\hat{S}\left(\xi_1, \xi_2\right)
 \end{equation}
where the squeezing transformation is defined by
\begin{equation}
\label{26}
\hat{S}(\xi_1, \xi_2)=\hat{S}_1(\xi_1)\hat{S}_2(\xi_2),
\end{equation}
with $\hat{S}_i(\xi_i)$ being the single-mode squeezing operator defined by
\begin{equation}
\label{27}
\hat{S}_i(\xi_i)=\exp\left[- \frac{\xi_i}{2}\hat{a}^{\dagger 2}_i
+\frac{\xi^*_i}{2}\hat{a}^{2}_i\right], \hspace{0.5cm} (i=1, 2).
\end{equation}

It is easy to show that when the squeezing parameters are chosen
as
\begin{equation}
\label{28}
\xi_i=\frac{1}{4}\ln\left(\frac{\omega'_i-(-1)^i\lambda}{\omega'_i+(-1)^i\lambda}\right),
\hspace{1cm} (i=1, 2),
\end{equation}
the squeezing transformation can diagonalize the Hamiltonian
$\hat{H}_2$ as the following form
\begin{equation}
\label{29}
\hat{H}_3=\hbar\Omega_1\hat{a}^{\dagger}_1\hat{a}_1 +
     \hbar\Omega_2\hat{a}^{\dagger}_2\hat{a}_2,
\end{equation}
where the frequencies $\Omega_i$ is given by
\begin{equation}
\label{30}
\Omega_i=\sqrt{\omega_i'^2-\lambda^2}=2e\sqrt{\frac{E_J}{C_{t_i} -
(-1)^iC_C}}, \hspace{0.5cm} (i=1,2)
\end{equation}

In the derivation of the Hamiltonian (\ref{26}) we have used the
following formula
\begin{eqnarray}
\label{32}
\hat{S}^{\dagger}_i(\xi_i)\hat{a}_i\hat{S}_i(\xi_i)&=&\hat{a}_i\cosh \xi_i - \hat{a}^{\dagger}_i \sinh \xi_i, \\
\label{33}
\hat{S}^{\dagger}_i(\xi_i)\hat{a}^{\dagger}_i\hat{S}_i(\xi_i)&=&
\hat{a}^{\dagger}_i \cosh \xi_i - \hat{a}_i \sinh \xi_i,
\end{eqnarray}

Assume that the two islands are initially in a coherent state
\begin{equation}
\label{34}
|\Psi(0)\rangle=\hat{D}(\alpha_1,\alpha_2)|0,0\rangle,
\end{equation}
where $\hat{D}(\alpha_1,\alpha_2)$ is the displacement operator
with respect to the two modes defined by Eq.
(\ref{19}).

%\begin{equation} \label{35}
%\hat{D}(\alpha_1,\alpha_2)=\prod^2_{j=1}\exp\left[\alpha_i\hat{a}^{\dagger}_i
%-\alpha^*_i\hat{a}_i\right].
%\end{equation}

After making three anti-transformations $\hat{S}^{\dagger}(\xi_1, \xi_2)$, $\hat{B}^{\dagger}(\pi/4)$,
and $\hat{D}^{\dagger}(\alpha_{01}, \alpha_{02})$, the initial state becomes
\begin{equation}
\label{36}
|\Psi(0)\rangle_3=\hat{S}^{\dagger}(\xi_1, \xi_2)\hat{B}^{\dagger}\left(\frac{\pi}{4}\right)
\hat{D}^{\dagger}(\alpha_{01}, \alpha_{02})|\alpha_1, \alpha_2\rangle.
\end{equation}

Then at an arbitrary time $t$, the wave function of the system is given by
\begin{equation}
\label{37}
|\Psi(t)\rangle_3=\exp\left(-\frac{i}{\hbar}\hat{H}_3t\right)
|\Psi(0)\rangle_3.
\end{equation}

Making use of the following formula
\begin{equation}
\label{38}
\hat{B}^{\dagger}\left(\frac{\pi}{4}\right)\hat{D}(\alpha_1-\alpha_{01},
\alpha_2-\alpha_{02})\hat{B}\left(\frac{\pi}{4}\right)=\hat{D}(\gamma_1, \gamma_2),
\end{equation}
where the two parameters $\gamma_1$ and $\gamma_2$ are defined as
\begin{eqnarray}
\label{39}
\gamma_i&=&\frac{1}{\sqrt{2}}\left[(\alpha_1+(-1)^i\alpha_2)-(\alpha_{01}+(-1)^i\alpha_{02})\right],
\end{eqnarray}
%&=&i\left(
%\frac{E_C}{8E_J}\right)^{\frac14}\left[(n_1'+(-1)^in_2')-
%(n_{g_1}+(-1)^in_{g_2})\right] \nonumber \\
%&+&\left(\frac{E_J}{32E_C}\right)^{\frac14}\left[\phi_1'+(-1)^i\phi_2'\right],
%\hspace{0.5cm} (i=1, 2),
%\end{eqnarray}
%where the initial number of excess cooper pairs on each island
%$n'_i$, and initial phase of each junction $\phi'_i$ are given by
%\begin{equation} n'_i=\left(
%\frac{8E_J}{E_C}\right)^{1/4}\textrm{Im}(\alpha_i),
%\hspace{0.3cm} \phi'_i=\left(
%\frac{32E_C}{E_J}\right)^{1/4}\textrm{Re}(\alpha_i),
%\end{equation}

We can express the evaluation of the state as
\begin{equation}
\label{41} |\Psi(t)\rangle_3=\prod^2_{j=1}\left[
\hat{R}_j\left(-i\Omega_jt\right)\hat{S}(-\xi_j)\hat{D}(\gamma_j)\right]|0,
0\rangle.
\end{equation}
which indicates that  when the system is initially in a coherent
state, the system evolves by a rotated-squeezed-coherent state in
the transformed representation with  a displacing transformation,
a beam-splitter-like transformation, and a squeezing
transformation. It should be mentioned that the state (\ref{41})
is an entangled state in the original representation, although it
is a non-entangled state in the transformed representation.

From (\ref{37}) we can obtain the number of Cooper pairs on the
$i$-th island
\begin{eqnarray}
\label{42} N_i&=& n_{g_1} + \frac{1}{2}\left[\rho_2 \sin (\Omega_2 t + \delta _2 )\right. \nonumber \\
& &\left.- (-1)^i\rho_1 \sin (\Omega _1 t + \delta _1 ) \right],
\hspace{0.5cm} (i=1, 2).
\end{eqnarray}
with the amplitudes and phase shifts given by
\begin{eqnarray}
\rho_i&=&\left({\frac{{E_C }}{{32E_J }}}
\right)^{\frac{1}{4}}\left[(\textrm{Im}\gamma_i)^2+e^{4\xi_i}(\textrm{Re}\gamma_i)^2
\right]^{1/2}, \\
\delta_i&=&\tan^{-1}\left(-e^{-2\xi_i}\tan\varphi_{\gamma_i}\right).
\end{eqnarray}

%\begin{eqnarray}
%\rho_i^2&=&\left[(n_1'+(-1)^in_2') - (n_{g_1}+(-1)^in_{g_2})
%\right]^2 \nonumber \\
%&&+ \frac{E_J}{2E_C - E_{12}}\left[\phi_1'+(-1)^i\phi_2'\right]^2
%\end{eqnarray}
Form Eq. (\ref{42}) it can be seen that the number of Cooper pairs
on each island oscillates periodically, and one can control
periods and amplitudes of these oscillations through changing
circuit parameters and gate voltages.

 The sum of Cooper pairs in the two islands is given by
\begin{eqnarray}
\label{43} \langle N_1+N_2\rangle
=(n_{g_1}+n_{g_2})+\rho_2\sin(\Omega_2t+\delta_2),
\end{eqnarray}
and the difference of Cooper pairs in the two islands is given by
\begin{eqnarray}
\label{44} \langle
N_1-N_2\rangle=(n_{g_1}-n_{g_2})+\rho_1\sin(\Omega_1t+\delta_1).
\end{eqnarray}

Making use of the wavefunction (\ref{41}), we find that quantum
fluctuation in each island is given by
\begin{eqnarray}
\label{45} \langle ( {\Delta \hat N_i } )^2  \rangle=\left(
{\frac{{E_J }}{{32E_C }}} \right)^{\frac12} \sum\limits_{j = 1}^2
[ \cos ^2 \Omega _j t + e^{4\xi _j } \sin ^2 \Omega _j t],
\end{eqnarray}
which implies that the quantum fluctuation of the Cooper pairs in
each island oscillates periodically with the time evolution. The
amplitude of the quantum fluctuation is dependent of the squeezing
parameter $\xi_i$, i.e., the Josephson coupling energy and
charging energy of the system under our consideration.

In summary, we have investigated quantum dynamics of two capacitively coupled superconducting
islands via Josephson junctions. We have shown that the system under our consideration can be reduced to
 two coupling harmonic oscillators. We have obtained an exact solution of the system when two harmonic
 oscillators are initially in coherent states, and found that the system evolves by a rotated-squeezed-coherent state.
 We have also studied the Cooper-pair evolution and quantum fluctuations in the two superconducting islands and found
 that the number of the Cooper pairs in the two islands evolves periodically.

%\acknowledgments

This work is supported by the National Fundamental Research
Program (2001CB309310), the National Natural Science Foundation,
the State Education Ministry of China, and the Innovation Funds
from Chinese Academy of Sciences via the Institute of Theoretical
Physics, Academia, Sinica.

\end{document}